\documentclass[review]{elsarticle}
\usepackage{amsfonts,bbm,amssymb,amsmath,slashed,mathrsfs,graphicx,tikz,color}
\usepackage{lineno,hyperref}

\newtheorem{theorem}{Theorem}
\newtheorem{lemma}{Lemma}
\newtheorem{corollary}[theorem]{Corollary}

\newdefinition{remark}{Remark}[section]
\newdefinition{definition}{Definition}[section]
\newdefinition{postulate}{Postulate}
\newproof{proof}{{\noindent\it Proof}}

\def\QEDopen{{\setlength{\fboxsep}{0pt}\setlength{\fboxrule}{0.2pt}\fbox{\rule[0pt]{0pt}{1.3ex}\rule[0pt]{1.3ex}{0pt}}}}
\def\QED{\QEDopen}

\def\Q.E.D{\hfill\QED}

\journal{}

\begin{document}

\begin{frontmatter}



\title{Model-Checking PCTL Properties of Stateless Probabilistic Pushdown Systems}

\author{Deren Lin}
\address{Xiamen City, China}

\author{Tianrong Lin\corref{ca}}
\address{National Hakka University,  China}
\cortext[ca]{Corresponding author.}

\begin{abstract}
In this short communication, we resolve a longstanding open question in the probabilistic verification of infinite-state systems. We show that model checking {\em stateless probabilistic pushdown systems (pBPA)} against {\em probabilistic computational tree logic (PCTL)} is generally undecidable.
\end{abstract}


\end{frontmatter}
\section{Introduction}
\label{sec:introduction}

{\em Model checking} \cite{CGP99} is an essential tool for formal verification, which is an interesting and important topic in the research field of logic in computer science and particularly plays an important role in verification of digital circuit (chips), in which one describes the system to be verified as a model of some logic, expresses the property to be verified as a formula in that logic, and then checks by using automated algorithms that the formula holds or not in that model, see e.g. the standard textbook \cite{BK08} by Baier et al. In particular, the famous work \cite{VW94} investigated extensions of temporal logic by connectives defined by finite automata on infinite words, which are important directions in model-checking. Traditionally, model checking has been applied to finite-state systems and non-probabilistic programs. During the last two decades, researchers have paid much attention to model-checking of probabilistic infinite-state systems, see e.g. \cite{EKM06}.  Apart from the above mentioned works, there are many other excellent works model-checking on infinite-state systems, such as \cite{BH05} where the {\em Well-structured transition systems (WSTS)} were investigated, \cite{QR05} in which context-bounded model checking of concurrent software was studied, and \cite{RHC05} in which the algorithms for model-checking CSL (continuous stochastic logic) against infinite-state continuous-time Markov chains are developed.

Among the probabilistic infinite-state systems, one is the {\em probabilistic pushdown systems}, which were dubbed ``{\em probabilistic pushdown automata}" in \cite{BBFK14,Bra07,EKM06}, the input alphabet of which contains only one symbol. Throughout the paper, such a limited version of probabilistic pushdown automata will be dubbed ``{\em probabilistic pushdown system}". Their model-checking question, initiated in \cite{EKM06}, has attracted a lot of attention, see e.g. \cite{Bra07,BBFK14}, in which the model-checking of {\em stateless probabilistic pushdown systems} (pBPA) against PCTL$^*$ was resloved. However, the question of model-checking of {\em stateless probabilistic pushdown systems} (pBPA) against PCTL still left open in \cite{Bra07,BBFK14}, which was first proposed in \cite{EKM06}.

The main goal of this paper is to provide a solution to the aforementioned longstanding open question. Our main purpose here is that we are willing to tackle an open question in the field to get a taste of this subject. Our main method for handling this question is based on the techniques of construction formulas presented in \cite{Bra07,BBFK14}, together with our own new observations and ideas. Namely, we try to construct PCTL formulas which encode the modified {\em Post Correspondence Problem} from our ideas. It should be pointed out that although we continue to employ some technique presented in \cite{Bra07, BBFK14}, our contributions are not only to be just solving a math question based on the already known techniques, because there are many new observation and idea hidden behind the solution. In addition, by the techniques presented in \cite{BBFK14, Bra07} alone, it seems impossible to answer this question, which means that it requires new angles of viewpoint (see e.g. Remark \ref{remark7} and Remark \ref{remark8}).

\begin{theorem}
\label{theorem1}
The model-checking of stateless probabilistic pushdown system (pBPA) against probabilistic computational tree logic PCTL is generally undecidable.
\end{theorem}

Because the class of stateless probabilistic pushdown systems is a sub-class of probabilistic pushdown systems, and the logic of PCTL is a sublogic of PCTL$^*$, by Theorem \ref{theorem1}, we also arrive at the undecidability results in \cite{BBFK14}. Namely, the following two corollaries:

\begin{corollary}
The model-checking of probabilistic pushdown systems (pPDS) against probabilistic computational tree logic PCTL is generally undecidable.\Q.E.D
\end{corollary}

\begin{corollary}
The model-checking of stateless probabilistic pushdown systems (pBPA) against probabilistic computational tree logic PCTL$^*$ is generally undecidable.\Q.E.D
\end{corollary}

The rest of this paper is structured as follows: in the Section \ref{sec:preliminaries}, some basic notions will be reviewed and useful notation will be fixed. Section \ref{sec:proof_of_theorem_1} is devoted to the proof of the main theorem, and the last Section is for conclusions.

\section{Preliminaries}
\label{sec:preliminaries}

For convenience, most notation in probabilistic verification will follow the papers \cite{Bra07,BBFK14}. For probability theory, the reader is referred to \cite{Shi95} by Shiryaev or \cite{Loe78a,Loe78b} by Lo\`{e}ve. Let $|A|$ denote the cardinality of any finite set $A$. Let $\Sigma$ and $\Gamma$ denote non-empty finite alphabets. Then $\Sigma^*$ is the set of all finite words (including empty word $\epsilon$) over $\Sigma$, and $\Sigma^+ = \Sigma^*\backslash \{\epsilon\}$. For any word $w\in \Sigma^*$, $|w|$ represents its length. For example, let $\Sigma = \{0, 1\}$, then $|\epsilon| = 0$ and $|001101| = 6$.

\subsection{Markov Chains}

Roughly, {\em Markov chains} are {\em probabilistic transition systems}, which are accepted as the most popular operational model for the evaluation of the performance and dependability of information-processing systems. For more details, see \cite{BK08}.

\begin{definition}
\label{definition1}
A {\em (discrete) Markov chain} is a triple $\mathcal{M}=(S,\delta,\mathcal{P})$ where $S$ is a finite or countably infinite set of states, $\delta\subseteq S\times S$ is a transition relation such that for each $s\in S$ there \textcolor{red}{exists} $t \in S$ such that $(s, t) \in\delta$, and $\mathcal{P}$ is a function from domain $\delta$ to range $(0, 1]$ which to each transition $(s, t) \in \delta$ assigns its probability $\mathcal{P}(s, t)$ such that $\sum_t\mathcal{P}(s, t) = 1$ for each $s \in S$.
\end{definition}

\begin{remark}
\label{remark1}
 $\sum_t\mathcal{P}(s,t)$ means $\mathcal{P}(s,t_1)+\mathcal{P}(s,t_2)+\cdots+\mathcal{P}(s,t_i)$ where $\{(s,t_1),(s,t_2),\cdots, (s,t_i)\}\subseteq\delta$ is the set of all transition relations whose current state is $s$.
\end{remark}

A path in $\mathcal{M}$ is a finite or infinite sequence of states of $S: \pi = s_0s_1\cdots\in S^n$ (or $\in S^{\omega}$) where $n\in\mathbb{N}_1=\{1,2,\cdots\}$ such that $(s_i, s_{i+1}) \in \delta$ for each $i$. A run of $\mathcal{M}$ is an infinite path. We denote the set of all runs in $\mathcal{M}$ by $Run$, and $Run(\pi')$ to denote the set of all runs starting with a given finite path $\pi'$. If a run $\pi$ starts with a given finite path $\pi'$, then we denote this case as $\pi'\in prefix(\pi)$. Let $\pi$ be a run, then $\pi[i]$ denotes the state $s_i$ of $\pi$, and $\pi_i$ the run $s_is_{i+1}\cdots$. In this way, it is clear that $\pi_0 = \pi$. Further, a state $s'$ is $reachable$ from a state $s$ if there is a {\em finite path} starting in $s$ and ending at $s'$.

For each $s \in S$, $(Run(s), \mathcal{F},\mathcal{P})$ is a probability space, where $\mathcal{F}$ is the $\sigma$-field generated by all {\em basic cylinders} $Cyl(\pi)$ and $\pi$ is a finite path initiating from $s$, $Cyl(\pi)=\{\widetilde{\pi}\in Run(s) : \pi\in prefix(\widetilde{\pi})\}$, and $\mathcal{P} : \mathcal{F} \rightarrow [0, 1]$ is the unique probability measure such that $\mathcal{P}(Cyl(\pi)) = \prod_{1\leq i\leq |\pi|-1} \mathcal{P}(s_i, s_{i+1})$ where $\pi = s_1s_2\cdots s_{|\pi|}$ and $s_1=s$.

\subsection{Probabilistic Computational Tree Logic}

The logic PCTL was originally introduced in \cite{HJ94}, where the corresponding model-checking question has been focused mainly on {\em finite-state Markov chains}.

Let $AP$ be a fixed set of atomic propositions. Formally, the syntax of {\em probabilistic computational tree logic} PCTL is given by
$$\aligned
 \Phi&::=p\text{ $|$ }\neg\Phi\text{ $|$ }\Phi_1\wedge\Phi_2\text{ $|$ }\mathcal{P}_{\bowtie r}(\varphi)\\
 \varphi&::={\bf X}\Phi\text{ $|$ } \Phi_1{\bf U}\Phi_2
\endaligned$$

where $\Phi$ and $\varphi$ denote the state formula and path formula, respectively; $p \in AP$ is an atomic proposition. In the above, $\bowtie$ is drawn from $\{>, =\}$\footnote{ The comparison relations such as ``$\geq$", ``$\leq$", and ``$<$" have been excluded, as ``$\geq$" and ``$=$" are sufficient enough for our discussion.},  $r$ is an rational with $0 \leq r \leq 1$.

Let $\mathcal{M} = (S, \delta,\mathcal{P})$ be a {\em Markov chain} and $\nu: S \rightarrow 2^{AP}$ an assignment and the symbol {\bf true} the abbreviation of always true. Then the semantics of PCTL, over $\mathcal{M}$, is given by the following rules
$$\aligned
\mathcal{M},s\models^{\nu}{\bf true}   \,\,\,&\text{     }\,\,\,\text{    for any $s\in S$}\\
\mathcal{M},s\models^{\nu}p                     \,\,\,&{\rm iff}\,\,\,\text{ $s\in\nu(p)$}\\
\mathcal{M},s\models^{\nu}\neg\Phi              \,\,\,&{\rm iff}\,\,\,\text{ $\mathcal{M},s\not\models^{\nu}\Phi$}\\
\mathcal{M},s\models^{\nu}\Phi_1\wedge \Phi_2   \,\,\,&{\rm iff}\,\,\,\text{ $\mathcal{M},s\models^{\nu}\Phi_1$}\\
&\qquad\text{ and $\mathcal{M},s\models^{\nu}\Phi_2$}\\
\mathcal{M},s\models^{\nu}\mathcal{P}_{\bowtie r}(\varphi)  \,\,\,&{\rm iff}\,\,\,\\
\text{ $\mathcal{P}$}&\text{$(\{\pi\in Run(s):\mathcal{M},\pi\models^{\nu}\varphi\})\bowtie r$}\\
\mathcal{M},\pi\models^{\nu}{\bf X}\Phi \,\,\,&{\rm iff}\,\,\,\text{$\mathcal{M},\pi[1]\models^{\nu}\Phi$}\\
\mathcal{M},\pi\models^{\nu}\Phi_1{\bf U}\Phi_2 \,\,\,&{\rm iff}\,\,\,\text{$\exists k\geq 0$  s.t. $\mathcal{M},\pi[k]\models^{\nu}\Phi_2$}\\
&\text{ and $\forall j. 0\leq j<k:\mathcal{M},\pi[j]\models^{\nu}\Phi_1$}
\endaligned$$

\begin{remark}
\label{remark2}
The logic PCTL$^*$ extends PCTL by deleting the requirement that any temporal operator must be proceeded by a state formula (Thus, the logic PCTL can be regarded as a sublogic of PCTL$^*$), and its path formulas are generated by the following syntax:
$$
\varphi::=\Phi\text{ $|$ }\neg\varphi\text{ $|$ }\varphi_1\wedge\varphi_2\text{ $|$ }{\bf X}\varphi\text{ $|$ }\varphi_1{\bf U}\varphi_2.
$$
\end{remark}

The difference between PCTL and PCTL$^*$ is very clear: a well-defined PCTL formula is definitely a well-defined PCTL$^*$ formula. However, the inverse is not necessarily true. The semantics of PCTL$^*$ path formulas over $\mathcal{M}$ are defined as follows:
$$\aligned
\mathcal{M},\pi\models^{\nu}\Phi\,\,\,&{\rm iff}\,\,\,\text{$\mathcal{M},\pi[0]\models^{\nu}\Phi$}\\
\mathcal{M},\pi\models^{\nu}\neg\varphi\,\,\,&{\rm iff}\,\,\,\text{$\mathcal{M},\pi\not\models^{\nu}\varphi$}\\
\mathcal{M},\pi\models^{\nu}\varphi_1\wedge\varphi_2\,\,\,&{\rm iff}\,\,\,\text{ $\mathcal{M},\pi\models^{\nu}\varphi_1$ and $\mathcal{M},\pi\models^{\nu}\varphi_2$}\\
\mathcal{M},\pi\models^{\nu}{\bf X}\varphi\,\,\,\,&{\rm iff}\,\,\,\mbox{$\mathcal{M},\pi_1\models^{\nu}\varphi$}\\
\mathcal{M},\pi\models^{\nu}\varphi_1{\bf U}\varphi_2\,\,\,&{\rm iff}\,\,\,\text{$\exists k\geq 0$ s.t. $\mathcal{M},\pi_k\models^{\nu}\varphi_2$}\\
&\,\,\,\text{ and $\forall j.0\leq j< k$: $\mathcal{M},\pi_j\models^{\nu}\varphi_1$}
\endaligned$$

\begin{remark}
\label{remark3}
The abbreviation of ``s.t." means ``such that". The logic PCTL or PCTL$^*$ can be interpreted over an {\em Markov decision process} (MDP) $\mathcal{M}$ in the similar way that we just did with the {\em Markov chain}. But it is outside our topic here.
\end{remark}

\subsection{Probabilistic Pushdown Systems}

Let us recall the definition of the {\em probabilistic pushdown systems}, being as follows:
\begin{definition}
\label{definition2}
A {\em probabilistic pushdown system (pPDS)} is a tuple $\Xi = (Q, \Gamma, \delta,\mathcal{P})$ where $Q$ is a finite set of control states, $\Gamma$ a finite stack alphabet, $\delta \subseteq (Q \times\Gamma) \times(Q\times\Gamma^*)$ a finite set of rules satisfying
\begin{itemize}
\item { each $(p,X)\in Q\times\Gamma$ \textcolor{red}{satisfying that} there is at least one rule of the form $((p,X),(q,\alpha))\in\delta$; In the following we will write $(p,X)\rightarrow(q,\alpha)$ instead of $((p,X),(q,\alpha))\in\delta$.
}
\item { $\mathcal{P}$ is a function from $\delta$ to $[0,1]$ which to each rule $(p,X)\rightarrow(q,\alpha)$ in $\delta$ assigns its probability $\mathcal{P}((p,X)\rightarrow(q,\alpha))\in[0,1]$ s.t. for each $(p,X)\in Q\times\Gamma$ satisfying that $\sum_{(q,\alpha)}\mathcal{P}((p,X)\rightarrow(q,\alpha))=1$. Furthermore, without loss of generality, we assume $|\alpha|\leq 2$. The configurations of $\triangle$ are elements in $Q\times\Gamma^*$.
}
\end{itemize}
\end{definition}

The {\em stateless probabilistic pushdown system (pPBA)} is a {\em probabilistic pushdown system (pPDS)} whose state set $Q$ is a singleton (or, we can just omit $Q$ without any influence).

\begin{definition}
\label{definition3}
A {\em stateless probabilistic pushdown system (\textcolor{red}{shortly, pBPA}\footnote{\textcolor{red}{Or,``B" stands for ``stateless".}})} is a triple $\triangle= (\Gamma, \delta,\mathcal{P})$, whose configurations are elements $\in\Gamma^*$, where $\Gamma$ is a finite stack alphabet, $\delta$ a finite set of rules satisfying
\begin{itemize}
\item {for each $X\in\Gamma$, there is at least one rule $(X,\alpha)\in\delta$ where $\alpha\in\Gamma^*$. In the following, we write $X\rightarrow\alpha$ instead of $(X,\alpha)\in\delta$;  We assume, w.l.o.g., that $|\alpha|\leq 2$.}
\item {$\mathcal{P}$ is a function from $\delta$ to $[0,1]$ which to every rule $X\rightarrow\alpha$ in $\delta$ assigns its probability $\mathcal{P}(X\rightarrow\alpha)\in[0,1]$ s.t. for each $X\in\Gamma$, it meets the condition that $\sum_{\alpha}\mathcal{P}(X\rightarrow\alpha)=1$.
     }
\end{itemize}
\end{definition}

Given a $pPDS$ or $pBPA$ $\triangle$, it induces an {\em infinite-state Markov chain} $\mathcal{M}_{\triangle}$. The model-checking question for properties expressed by the PCTL formula $\Psi$ is defined to determine whether $\mathcal{M}_{\triangle}\models^{\nu}\Psi$.

As shown in \cite{EKS03}, if there are no effective valuation assumptions, undecidable properties can be easily encoded to pushdown configurations. Thus, throughout the paper, we consider the same assignment as in \cite{EKS03,EKM06,BBFK14,Bra07}, which was called a regular assignment. More precisely, let $\triangle = (Q,\Gamma, \delta,\mathcal{P})$ be a {\em probabilistic pushdown system}, an assignment $\nu : AP \rightarrow 2^{Q\times\Gamma^*}$ ($2^{\Gamma^*}$ for a pBPA\footnote{\textcolor{red}{Since there is only one element in $Q$, thus we can explicitly omit the $Q$ due to that the configurations $(q,\gamma)$ and $(\gamma)$ are equivalent where $q\in Q=\{q\}$ and $\gamma\in\Gamma^*$.}}) is regular if $\nu(p)$ is a regular set for each $p \in AP$. In other words, {\em finite automata} $\mathcal{A}_p$ recognizes $\nu(p)$ over the alphabet $Q\cup\Gamma$, and $\mathcal{A}_p$ reads the stack of $\triangle$ from bottom to top. Furthermore, the regular assignment $\nu$ is simple if for each $p \in AP$ there is a subset of heads $H_p\subseteq Q\cup(Q\times\Gamma)$ such that $(q,\gamma\alpha)\in\nu(p)\Leftrightarrow(q,\gamma)\in H_p$, see e.g. \cite{BBFK14} for more details.

\subsection{Post Correspondence Problem}
\label{sec:post_correspondence_problem}

The {\em Post Correspondence Problem} (PCP), originally introduced and shown to be undecidable by Post \cite{Pos46}, has been used to show that many problems arising from formal languages are undecidable.

Formally, a PCP instance consists of a finite alphabet $\Sigma$ and a finite set $\{(u_i,v_i)\,|\,1\leq i\leq n\}\subseteq\Sigma^*\times\Sigma^*$ of $n$ pairs of strings over $\Sigma$, determining whether there is a word $j_1j_2\cdots j_k\in\{1,2,\cdots,n\}^+$ such that $u_{j_1}u_{j_2}\cdots u_{j_k}=v_{j_1}v_{j_2}\cdots v_{j_k}$.

There are numerous variants of the PCP definition, but the modified PCP \cite{BBFK14,Bra07} is the most convenient for our discussion in this paper. Since the word $w\in\Sigma^*$ is of finite length,\footnote{ See acknowledgements Section.} we can suppose that $m=\max\{|u_i|,|v_i|\}_{1\leq i\leq n}$. \textcolor{red}{If we put `$\bullet$' in the gap between two letters} of $u_i$ or $v_i$, to form the $u'_i$ or $v'_i$, such that $|u'_i|=|v'_i|=m$, then the modified PCP problem is to ask whether there exists  $j_1\cdots j_k\in\{1,\cdots,n\}^+$ such that the equation $u'_{j_1}\cdots u'_{j_k}=v'_{j_1}\cdots v'_{j_k}$ holds after erasing all `$\bullet$' in $u'_i$ and $v'_i$.

\begin{remark}
\label{remark4}
Essentially, the modified PCP problem is equivalent to the original PCP problem. That we stuff the $n$-pair strings $u_i$ and $v_i$ with `$\bullet$' to make them the same length is useful in Section \ref{sec:proof_of_theorem_1} to prove our main result.
\end{remark}

\section{Proof of Theorem \ref{theorem1}}
\label{sec:proof_of_theorem_1}

We are now proceeding to prove Theorem \ref{theorem1}. Throughout this section, we fix $\Sigma = \{A, B, \bullet\}$ and the stack alphabet $\Gamma$ of a pBPA is as follows:
$$\aligned
\Gamma=\{Z,Z',C,F,S,N,(x,y),&X_{(x,y)},G_i^j\,|\,(x,y)\in\Sigma\times\Sigma,\\
&1\leq i\leq n,1\leq j\leq m+1\}
\endaligned$$

The elements in $\Gamma$ also serve as symbols of atomic propositions whose senses will be clear from the following context. We will detail how to build the desirable {\em stateless probabilistic pushdown system} $\triangle=(\Gamma,\delta,\mathcal{P})$.

Similar to \cite{Bra07,BBFK14}, our $pBPA$ $\triangle$ also works in two steps, the first of which is to guess a possible solution to a modified PCP instance by storing pairs of words $(u_i,v_i)$ in the stack, which is done by the following transition rules (the probabilities of which are uniformly distributed):
\begin{equation}
  \label{eq1}
  \begin{split}
 Z\quad&\rightarrow\quad G_1^1Z'|\cdots|G_n^1Z';\\
  G_i^j\quad&\rightarrow\quad G_i^{j+1}(u_i(j),v_i(j));\\
   G_i^{m+1}\quad&\rightarrow\quad C|G_1^1|\cdots |G_n^1.
   \end{split}
\end{equation}

Obviously, we should let the symbol $Z$ serve as the initial stack symbol. It begins with pushing $G_i^1Z'$ ($\in\Gamma^*$) into the stack with probability $\frac{1}{n}$. Then, the symbol at the top of the stack is $G_i^1$ (we read the stack from left to right). The rules in (\ref{eq1}) state that \textcolor{red}{}$G_i^1$ is replaced with probability $1$ by $G_i^2(u_i(1),v_i(1))$. This process will be repeated until $G_i^{m+1}(u_i(m),v_i(m))$ is stored at the top of the stack, indicating that the first pair of $(u_i,v_i)$ has been stored.

Then, with the probability $\frac{1}{n+1}$, the $\triangle$ will go to push symbol $C$ or $G_i^1$ into the stack, depending on whether the guessing procedure is at the end or not. When the rule $G_i^{m+1}\rightarrow C$ is applied, the $\triangle$ goes to check whether the pairs of words stored in the stack are a solution of a modified PCP instance. It is clear that the above guess procedure will lead to a word $j_1j_2\cdots j_k\in\{1,2,\cdots ,n\}^+$ corresponding to the sequence of the words $(u_{j_1},v_{j_1}),(u_{j_2},v_{j_2}),\cdots,(u_{j_k},v_{j_k})$ pushed orderly into the stack. In addition, there are no other transition rules in the guessing-step for $\triangle$ except those illustrated by (\ref{eq1}). From the above explanation, we readily have the following:

\begin{lemma}[cf. \cite{BBFK14}, Lemma 3.2]
\label{lemma1}
  A configuration of the form $C\alpha Z'$ is reachable from $Z$ if and only if $\alpha\equiv(x_1,y_1)\cdots(x_l,y_l)$ where $x_j,y_j\in\Sigma$, and there is a word $j_1j_2\cdots j_k\in\{1,2,\cdots,n\}^+$ such that $x_l\cdots x_1=u_{j_1}\cdots u_{j_k}$ and $y_l\cdots y_1 = v_{j_1}\cdots v_{j_k}$. And the probability from $Z$ to $C\alpha Z'$ is $>0$. \Q.E.D
\end{lemma}

The next step is for $\triangle$ to verify a stored pair of words. Of course, this step should \textcolor{red}{be} slightly different from the one presented in \cite{Bra07,BBFK14} for us to construct a suitable PCTL formula describing this procedure, and the transition rules (the probabilities of them are uniformly distributed) are given as follows:
\begin{equation}
 \label{eq2}
  \begin{split}
  C\quad&\rightarrow\quad N,\\
  N\quad&\rightarrow\quad F\text{ $|$ }S,\\
   F\quad&\rightarrow\quad\epsilon,\\
    S\quad&\rightarrow\quad\epsilon,\\
     (x,y)\quad&\rightarrow\quad X_{(x,y)}\text{ $|$ }\epsilon,\\
      Z'\quad&\rightarrow\quad X_{(A,B)}\text{ $|$ }X_{(B,A)},\\
       X_{(x,y)}\quad&\rightarrow\quad\qquad\epsilon.
  \end{split}
\end{equation}

\begin{remark}
\label{remark5}
 We emphasize that, aside from the rules described in (\ref{eq2}), there are no other rules in the verifying-step for $\triangle$. In comparison to \cite{Bra07,BBFK14}, we have added another symbol $N$ to the stack alphabet $\Gamma$, which is for the purpose of using it for constructing a path formula starting with  {\bf X}.
\end{remark}

When the stack symbol $C$ is at the top of the stack, the $\triangle$ will check to see if the previous guess is a solution to the modified PCP instance. It first replaces $C$ with $N$ at the top of the stack, with a probability $1$, and then continues to push $F$ or $S$ into the stack, with a probability $\frac{1}{2}$, depending on whether the $\triangle$ wants to check $u$'s or $v$'s.

The following auxiliary Lemma is an adaptation from the Lemma 4.4.8 in \cite{Bra07}.
\begin{lemma}
\label{lemma2}
Let $\vartheta$ and $\overline{\vartheta}$ be two functions from $\{A,B,Z'\}$ to $\{0,1\}$, given by
$$
             \vartheta(X)=\left\{
                              \begin{array}{ll}
                                1, & \hbox{$X=Z'$;} \\
                                1, & \hbox{$X=A$;} \\
                                0, & \hbox{$X=B$.}
                              \end{array}
                            \right.\,\,\,
             \overline{\vartheta}(X)=\left\{
                                         \begin{array}{ll}
                                           1, & \hbox{$X=Z'$;} \\
                                           0, & \hbox{$X=A$;} \\
                                           1, & \hbox{$X=B$.}
                                         \end{array}
                                       \right.
$$

Further, let $\rho$ and $\overline{\rho}$ be two functions from $\{A,B\}^+Z'$ to $[0,1]$, given by
$$
 \rho(x_1x_2\cdots x_n)\overset{\mathrm{def}}{=}\sum_{i=1}^n\vartheta(x_i)\frac{1}{2^i},\quad
 \overline{\rho}(x_1x_2\cdots x_n)\overset{\mathrm{def}}{=}\sum_{i=1}^n\overline{\vartheta}(x_i)\frac{1}{2^i}.
$$

Then, for any $(u'_{j_1},v'_{j_1}),(u'_{j_2},v'_{j_2}),\cdots,(u'_{j_k},v'_{j_k})\in\{A,B\}^+\times\{A,B\}^+$,
\begin{equation}
 \label{eq3}
  \begin{split}
   u'_{j_1}u'_{j_2}\cdots u'_{j_k} = v'_{j_1}v'_{j_2}\cdots v'_{j_k}
           \end{split}
\end{equation}
if and only if
\begin{equation}
 \label{eq4}
  \begin{split}
  \rho(u'_{j_1}\cdots u'_{j_k}Z')+\overline{\rho}(v'_{j_1}v'_{j_2}\cdots v'_{j_k}Z') = 1
  \end{split}
\end{equation}
\end{lemma}
\begin{proof}
The ``only if" part is clear. Suppose that (\ref{eq3}) holds and that $u'_{j_1}\cdots u'_{j_k}=y_1\cdots y_l=v'_{j_1}\cdots v'_{j_k}$.
Then we have
$$\aligned
  &\rho(y_1\cdots y_lZ')+\overline{\rho}(y_1\cdots y_lZ')\\
   = &\sum_{i=1}^l(\vartheta(y_i)+\overline{\vartheta}(y_i))\frac{1}{2^i} + (\vartheta(Z')+\overline{\vartheta}(Z')\frac{1}{2^{l+1}}\\
              =&\sum_{i=1}^l\frac{1}{2^i} + \frac{2}{2^{l+1}}=1\quad\text{( by $\vartheta(x)+\overline{\vartheta}(x)=1$, $\forall x\in\{A,B\}$ )}
\endaligned$$

The ``if" part. If (\ref{eq4}) \textcolor{red}{holds true}, then (\ref{eq3}) must hold. Otherwise, suppose that $u'_{j_1}\cdots u'_{j_k}=x_1x_2\cdots x_l$ and $v'_{j_1}\cdots v'_{j_k}=y_1y_2\cdots y_m$ with $x_1x_2\cdots x_l\neq y_1y_2\cdots y_m$, then the result of $\rho(u'_{j_1}\cdots u'_{j_k}Z')+\overline{\rho}(v'_{j_1}\cdots v'_{j_k}Z') \neq 1$ deduced, which contradicts to (\ref{eq4}). Thus, the proof is complete.
\end{proof}

By Lemma \ref{lemma2}, if there exist two path formulas $\varphi_1$ and $\varphi_2$ to adhere to the probabilities of $\rho(u'_{j_1}\cdots u'_{j_k}Z')$ and $\overline{\rho}(v'_{j_1}\cdots v'_{j_k}Z')$, respectively, then we can successfully reduce the modified PCP problem to the model-checking question of whether $C\alpha Z\models^{\nu}\mbox{{\bf X}}(\mathcal{P}_{=t_1}(\varphi_1)\wedge \mathcal{P}_{=t_2}(\varphi_2))$ where the rationals $t_1$ and $t_2$ are s.t. $t_1+t_2=1$, which will be demonstrated by the following:

\begin{lemma}
\label{lemma3}
Let $\alpha = (u_{j_1},v_{j_1})(u_{j_2},v_{j_2})\cdots(u_{j_k},v_{j_k})\in\Sigma^*\times\Sigma^*$ be the pair of words pushed into the stack by $\triangle$. Let $(u'_i,v'_i)$, $1\leq i\leq j_k$, be the pair of words after erasing all $\bullet$ in $u_i$ and $v_i$. \textcolor{red}{Assume} $\varphi_1$ and $\varphi_2$ (defined later) be two path formulas satisfying the following
$$\aligned
\mathcal{P}(\{\pi\in\mbox{$Run(F\alpha Z')$}\,|\,\pi\models^{\nu}\varphi_1\})=&\rho(u'_{j_1}u'_{j_2}\cdots u'_{j_k}Z')\\
            \mathcal{P}(\{\pi\in\mbox{$Run(S\alpha Z')$}\,|\,\pi\models^{\nu}\varphi_2\})=&\overline{\rho}(v'_{j_1}v'_{j_2}\cdots v'_{j_k}Z').
\endaligned$$

Then
\begin{equation}
\label{eq5}
\begin{split}
 u'_{j_1}\cdots u'_{j_k} = v'_{j_1}\cdots v'_{j_k}
\end{split}
\end{equation}
if and only if $\mathcal{M}_{\triangle}, N\alpha Z'\models^{\nu}\mathcal{P}_{=\frac{t}{2}}(\varphi_1)\wedge\mathcal{P}_{\frac{1-t}{2}}(\varphi_2)$ where $t$: $0<t<1$ is a rational constant.
\end{lemma}

\begin{proof}
\textcolor{red}{First note that $t$ should not be considered as a free variable and cannot be $0$ or $1$.}

It is obvious that when $\alpha$ is pushed into the stack of $\triangle$, the stack's content is $C\alpha Z'$ (read from left to right). Note that there is only one rule, $C\rightarrow N$ which is applicable. Thus, with probability $1$, the content of the stack changes to $N\alpha Z'$.

The ``if" part. Suppose that $\mathcal{M}_{\triangle},N\alpha Z'\models^{\nu}\mathcal{P}_{=\frac{t}{2}}(\varphi_1)\wedge\mathcal{P}_{=\frac{1-t}{2}}(\varphi_2)$.

The probability of paths from $N$ that satisfy $\varphi_1$ is then $\frac{t}{2}$, and the probability of paths from $N$ that satisfy $\varphi_2$ is $\frac{1-t}{2}$. As a result, the probability of paths from $F$ satisfying $\varphi_1$ is $t$, while the probability of paths from $S$ satisfying $\varphi_2$ is $1 - t$. Because $\mathcal{P}(N\rightarrow F) = \mathcal{P}(N \rightarrow S) = \frac{1}{2}$, we have the following:
\begin{equation}
 \label{eq6}
 \begin{split}
 \rho(u'_{j_1}\cdots u'_{j_k}Z')+\overline{\rho}(v'_{j_1}\cdots v'_{j_k}Z')=t+(1-t) =1.
 \end{split}
\end{equation}
By (\ref{eq6}) and Lemma \ref{lemma2}, we conclude that (\ref{eq5}) holds true.

The ``only if" part. Obviously, that (\ref{eq5}) is true leads to $\rho(u'_{j_1}\cdots u'_{j_k}Z')+\overline{\rho}(v'_{j_1}\cdots v'_{j_k}Z') =1 \Rightarrow \rho(u'_{j_1}\cdots u'_{j_k}Z')=1-\overline{\rho}(v'_{j_1}\cdots v'_{j_k}Z')$. Namely, $\mathcal{P}(F\alpha Z'\models^{\nu}\varphi_1)=1-\mathcal{P}(S\alpha Z'\models^{\nu}\varphi_2)$. This together with $\mathcal{P}(N\rightarrow F)=\mathcal{P}(N\rightarrow S)=\frac{1}{2}$, further implies that $\mathcal{M}_{\triangle},N\alpha Z' \models^{\nu}\mathcal{P}_{=\frac{t}{2}}(\varphi_1)\wedge\mathcal{P}_{=\frac{1-t}{2}}(\varphi_2)$. The lemma follows.
\end{proof}

Now let us take $\alpha=(A,A)(A,\bullet)(\bullet,A)(B,B)$  as an example to see how to fix the path formulas $\varphi_1$ and $\varphi_2$, whose evolutionary process \footnote{ When it reaches the head of $X_{(A,z)}$ or $X_{(z,B)}$ where $z\in\Sigma$, we do not unfold the tree by (\ref{eq2}) any more.} is shown by the Figure \ref{fig:computational tree1} below:

\begin{figure}[htb]
\center{\includegraphics[width=6cm]{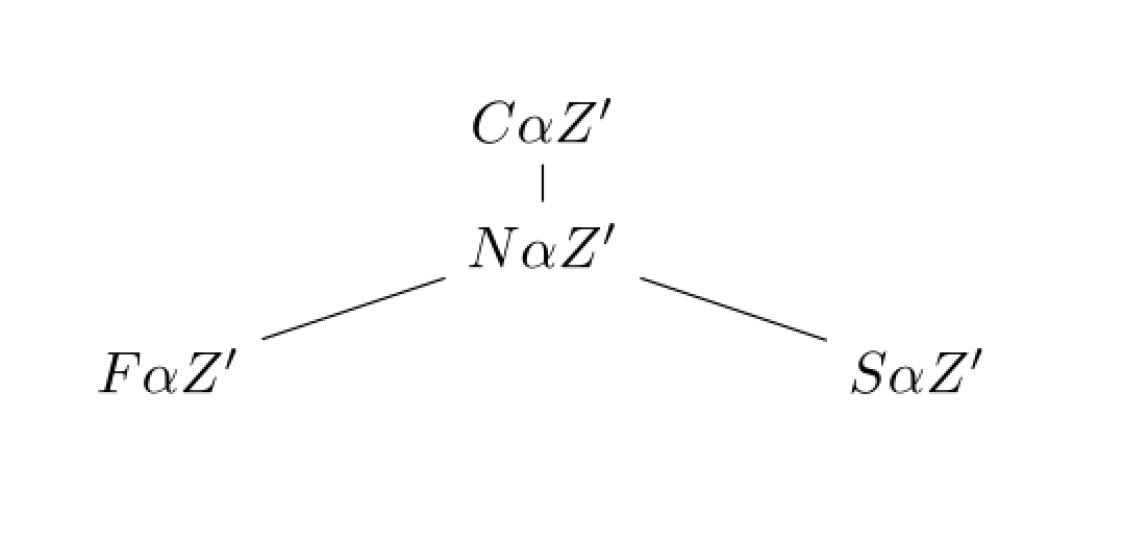}}
\caption{\label{fig:computational tree1} $C\alpha Z'$'s unfolding tree}
\end{figure}
  
and the Figure \ref{fig:computational tree2} below:
\begin{figure}[htb]
\center{\includegraphics[width=8cm]{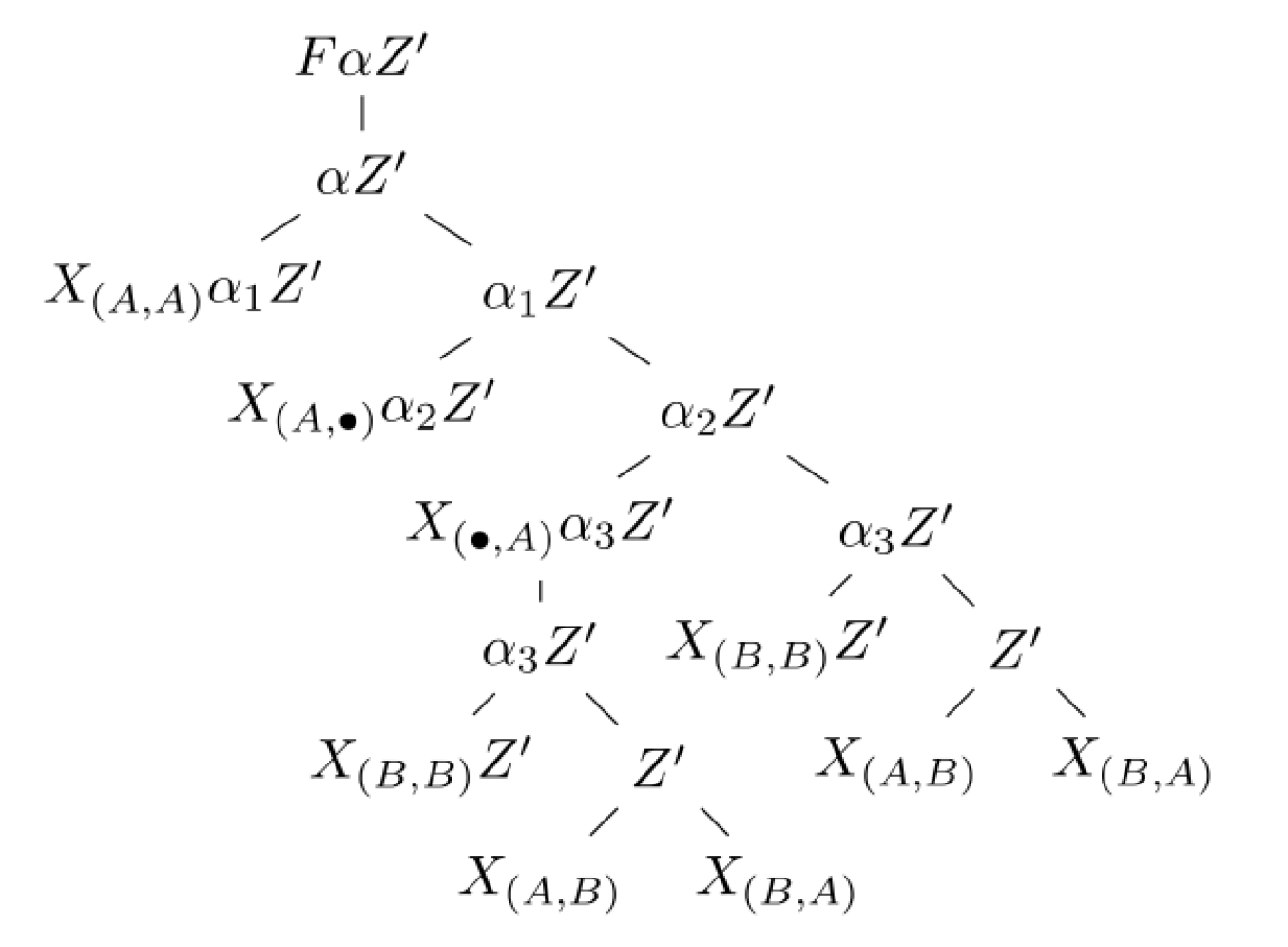}}
\caption{\label{fig:computational tree2} $F\alpha Z'$'s unfolding tree}
\end{figure}\\
where $\alpha_1=(A,\bullet)(\bullet,A)(B,B), \alpha_2=(\bullet,A)(B,B)$ and $\alpha_3=(B,B)$.

There are $4$ paths from state $F\alpha Z'$ to states that begin with $X_{(A,z)}$ where $z\in\Sigma$
$$\aligned
F\alpha Z'\rightarrow^{1}& \alpha Z'\rightarrow^{\frac{1}{2}} X_{(A,A)}\alpha_1Z'\\
&\text{(with probability $1\times \frac{1}{2}$)}\\
F\alpha Z'\rightarrow^{1}&\alpha Z'\rightarrow^{\frac{1}{2}}\alpha_1 Z'\rightarrow^{\frac{1}{2}}X_{(A,\bullet)}\alpha_2Z'\\
&\text{(with probability $1\times\frac{1}{2^2}$)}\\
 F\alpha Z'\rightarrow^1&\alpha Z\rightarrow^{\frac{1}{2}}\alpha_1 Z'\rightarrow^{\frac{1}{2}}\alpha_2Z'\\
 \rightarrow^{\frac{1}{2}}&X_{(\bullet,A)}\alpha_3Z'\rightarrow^1
 \alpha_3Z'\rightarrow^{\frac{1}{2}}Z'\\
 \rightarrow^{\frac{1}{2}}&X_{(A,B)}\\
 &\text{(with probability $1\times\frac{1}{2^5}$)}\\
  F\alpha Z'\rightarrow^1&\alpha Z\rightarrow^{\frac{1}{2}}\alpha_1 Z'\rightarrow^{\frac{1}{2}}\alpha_2Z'\\
  \rightarrow^{\frac{1}{2}}&\alpha_3Z\rightarrow^{\frac{1}{2}}Z'\rightarrow^{\frac{1}{2}}X_{(A,B)}\\
  &\text{(with probability $1\times\frac{1}{2^5}$)}
 \endaligned$$

So the total probability is $\frac{1}{2}+\frac{1}{2^2}+\frac{1}{2^4}$ which matches the value:
$$\aligned
\rho(AABZ')=&\vartheta(A)\frac{1}{2}+\vartheta(A)\frac{1}{2^2}+\vartheta(B)\frac{1}{2^3}+\vartheta(Z')\frac{1}{2^4}\\
=&\frac{1}{2}+\frac{1}{2^2}+\frac{1}{2^4}.
\endaligned$$

Observe that along the above paths, the states have no $S$ and no $X_{(B,z)}$ where $z\in\Sigma$ as their heads, and that we do not unfold the state with a head of $X_{(A,z)}$ any more. Thus, the above paths can be described by the following path formula:
\begin{equation}
\label{eq7}
\begin{split}
 \varphi_1\overset{\mathrm{def}}{=}(\neg S\wedge\bigwedge_{z\in\Sigma}\neg X_{(B,z)}){\bf U}(\bigvee_{z\in\Sigma}X_{(A,z)}).
\end{split}
\end{equation}

Similarly, we can obtain the path formula $\varphi_2$:
\begin{equation}
\label{eq8}
\begin{split}
 \varphi_2\overset{\mathrm{def}}{=}(\neg F\wedge\bigwedge_{z\in\Sigma}\neg X_{(z,A)}){\bf U}(\bigvee_{z\in\Sigma}X_{(z,B)}),
\end{split}
\end{equation}
of which the total probability along with the paths starting with $S\alpha Z'$ and ending in states that begin with $X_{(z,B)}$ where $z\in\Sigma$ matches the value:
$$\aligned
\overline{\rho}(AABZ')=&\overline{\vartheta}(A)\frac{1}{2}+\overline{\vartheta}(A)\frac{1}{2^2}+\overline{\vartheta}(B)\frac{1}{2^3}+\overline{\vartheta}(Z')\frac{1}{2^4}\\
 =&0\times\frac{1}{2}+0\times\frac{1}{2^2}+\frac{1}{2^3}+\frac{1}{2^4}.
\endaligned$$

\begin{remark}
\label{remark6}
In fact, the above two path formulas, $\varphi_1$ and $\varphi_2$, were used in \cite{Bra07} to specify the same paths illustrated above. Note that the atomic propositions $F$, $S$ and $X_{(z,z')}$ ($z,z'\in\Sigma$) are valid in exactly all configurations with the corresponding head, respectively. The reader can easily check that any path $\pi$ counted above satisfies the following
$$\aligned
\pi(k)&\models^{\nu} X_{(A,z)}\quad\mbox{(\textcolor{red}{for some} $k\geq 0$)}\\
\pi(i)&\models^{\nu} \neg S\wedge\bigwedge_{z\in\Sigma}\neg X_{(B,z)}\quad\text{(\textcolor{red}{for all }$0\leq i<k$)}.
\endaligned$$
\end{remark}

We will summarize the above analysis in the following Lemma, which establishes the connection between $\mathcal{P}(\{\pi\in\mbox{$Run(F\alpha Z')$}\text{ $|$ }\pi\models^{\nu}\varphi_1\})$ and the function $\rho$, and that between $\mathcal{P}(\{\pi\in\mbox{$Run(S\alpha Z')$}\text{ $|$ }\pi\models^{\nu}\varphi_2\})$ and the function $\overline{\rho}$, respectively. To prove it, we need to fix an additional notation: Let ${\rm trim}(b_1b_2\cdots b_n)$ denote the resultant word $\in\{A,B\}^*$ in which all the `$\bullet$' in $b_1b_2\cdots b_n$ are erased. Then ${\rm trim}(b_2b_3\cdots b_n)$ means the resultant word $\in\{A,B\}^*$ in which all the `$\bullet$' in $b_2b_3\cdots b_n$ are erased.

\begin{lemma}[cf. \cite{Bra07}]
\label{lemma4}
Let $\alpha = (x_1,y_1)(x_2,y_2)\cdots (x_l,y_l)\in\Sigma^*\times\Sigma^*$ be the pair of words pushed into the stack by $\triangle$, where $x_i,y_i\in\Sigma$, and $(u'_{j_i},v'_{j_i})$, $1\leq i\leq k$, the pair of words after erasing all $\bullet$ in $x_1x_2\cdots x_l$ and $y_1y_2\cdots y_l$. Then $\mathcal{P}(\{\pi\in\mbox{$Run(F\alpha Z')$}\,|\,\pi\models^{\nu}\varphi_1\})=\rho(u'_{j_1}u'_{j_2}\cdots u'_{j_k}Z')$ and $\mathcal{P}(\{\pi\in\mbox{$Run(S\alpha Z')$}\,|\,\pi\models^{\nu}\varphi_2\})=\overline{\rho}(v'_{j_1}v'_{j_2}\cdots v'_{j_k}Z')$.
\end{lemma}
\begin{proof}
We will show by induction on $l$ that $\mathcal{P}(F\alpha Z',\varphi_1)\overset{\rm def}{=}\mathcal{P}(\{\pi\in\text{$Run(F\alpha Z')$}\,|\,\pi\models^{\nu}\varphi_1\})=\rho({\rm trim}(x_1x_2\cdots x_l)Z')$; Similar arguments apply for
$\mathcal{P}(S\alpha Z',\varphi_2)\overset{\rm def}{=}\mathcal{P}(\{\pi\in\text{$Run(S\alpha Z')$}\,|\,\pi\models^{\nu}\varphi_2\})=\overline{\rho}({\rm trim}(y_1y_2\cdots y_l)Z')$.

Note that by (\ref{eq2}), $F\alpha Z'\rightarrow\alpha Z'$ with transition probability $1$, we have $\mathcal{P}(F\alpha Z',\varphi_1)=\mathcal{P}(\alpha Z',\varphi_1)$. Thus, to prove the lemma, we need only to show $\mathcal{P}(\alpha Z',\varphi_1)=\rho({\rm trim}(x_1x_2\cdots x_l)Z')$.
We give a proof by induction on $l$.\\
Base case: The case of $l=1$:
\begin{enumerate}
\item {if $x_1=\bullet$, then $\mathcal{P}((\bullet,z) Z',\varphi_1)=\frac{1}{2}\times\frac{1}{2}+\frac{1}{2}\times\frac{1}{2}=\frac{1}{2}=\rho({\rm trim}(\bullet)Z')$; see Figure \ref{fig:fig3} below:
    \begin{figure}[htb]
\center{\includegraphics[width=4.8cm]{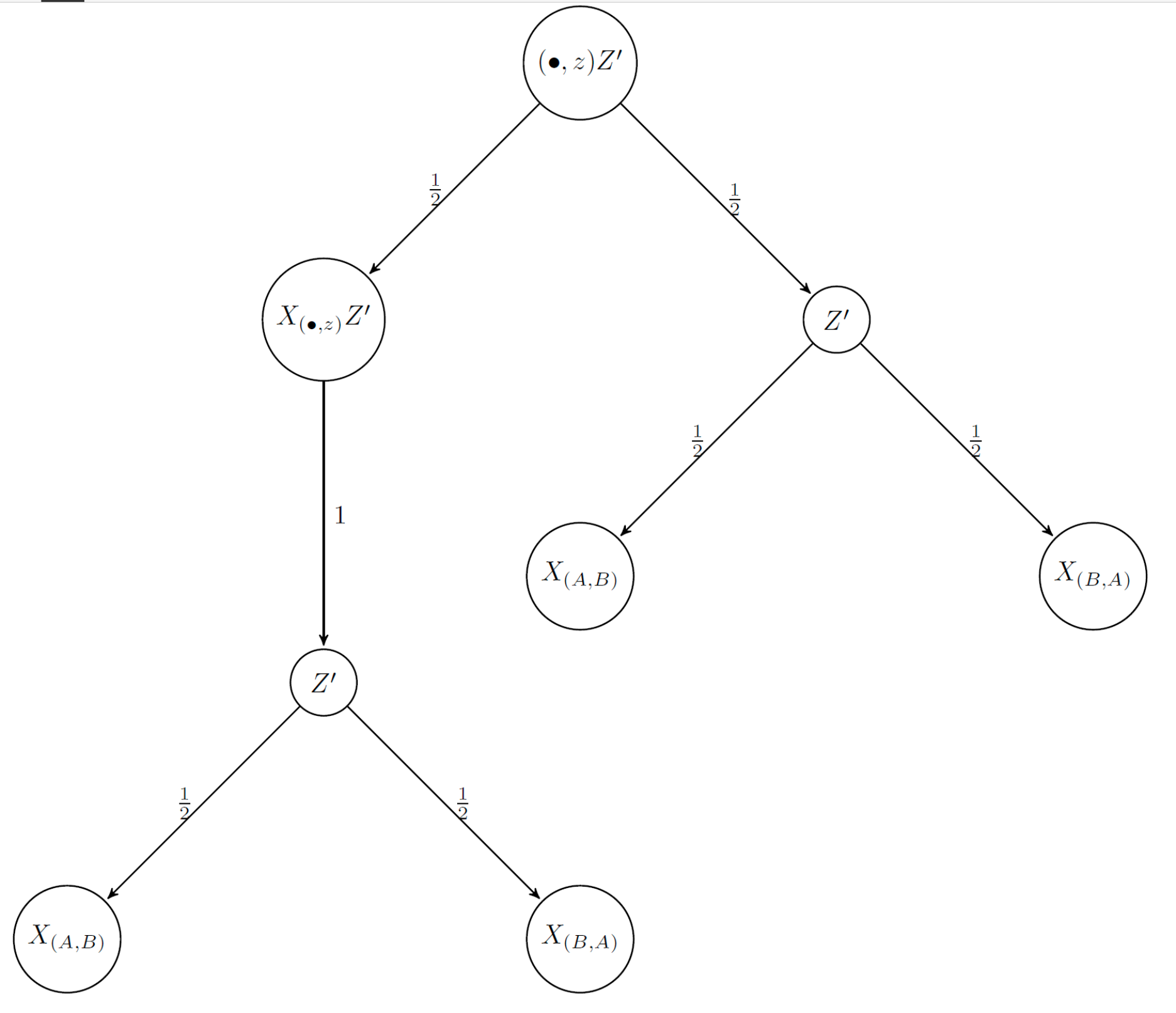}}
\caption{\label{fig:fig3} $(\bullet,z) Z'$'s unfolding tree}
\end{figure}}
\item {if $x_1=B$, then $\mathcal{P}((B,z)Z',\varphi_1)=\frac{1}{2}\times\frac{1}{2}=\frac{1}{2^2}=\rho({\rm trim}(B)Z')$; see Figure \ref{fig:fig4} above:
    \begin{figure}[htb]
\center{\includegraphics[width=4.8cm]{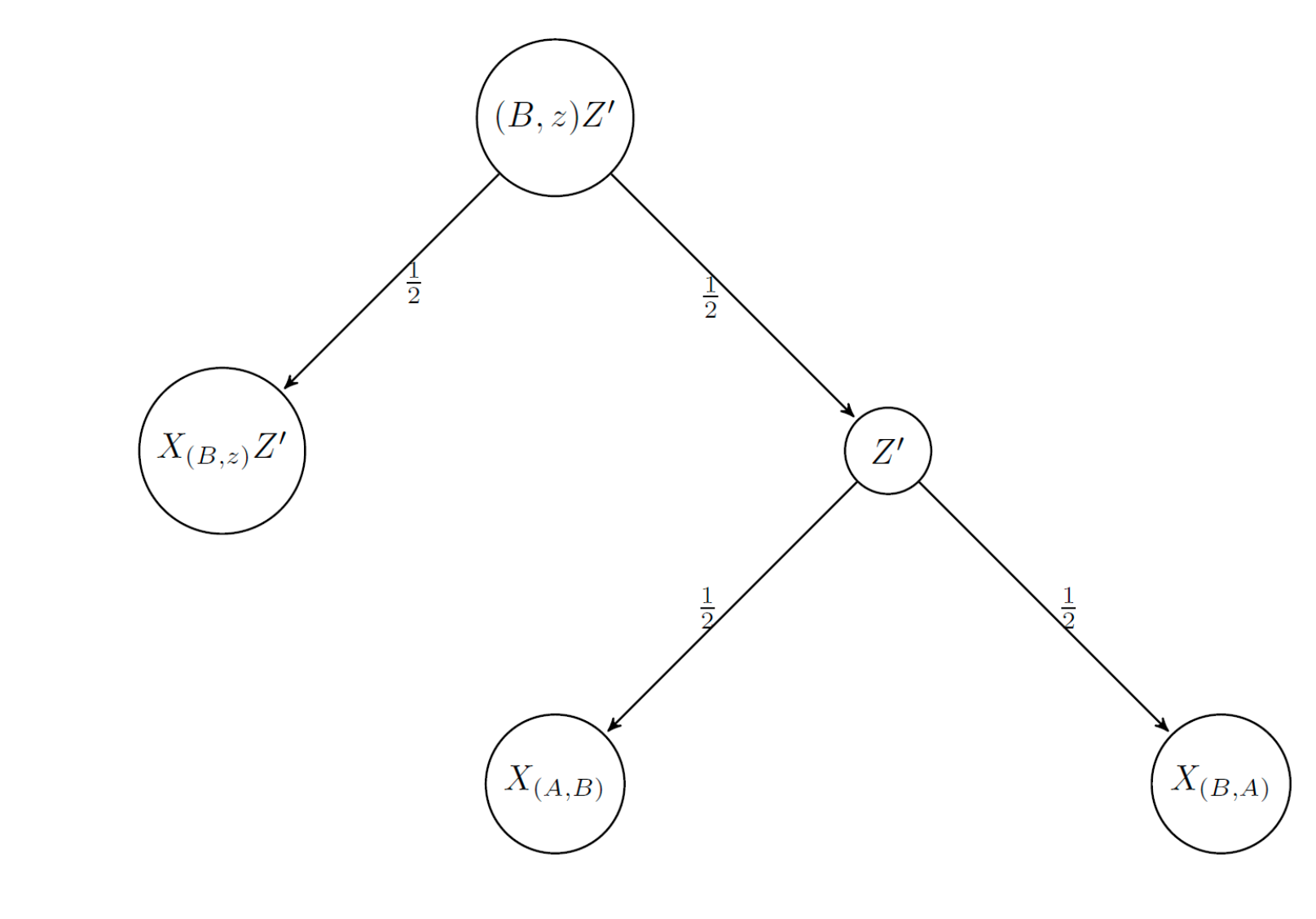}}
\caption{\label{fig:fig4} $(B,z)Z'$'s unfolding tree}
\end{figure}
    }
\item {if $x_1=A$, then $\mathcal{P}((A,z)Z',\varphi_1)=\frac{1}{2}+\frac{1}{2}\times\frac{1}{2}=\rho({\rm trim}(A)Z')$. See Figure \ref{fig:fig5} above:
    \begin{figure}[htb]
\center{\includegraphics[width=4.8cm]{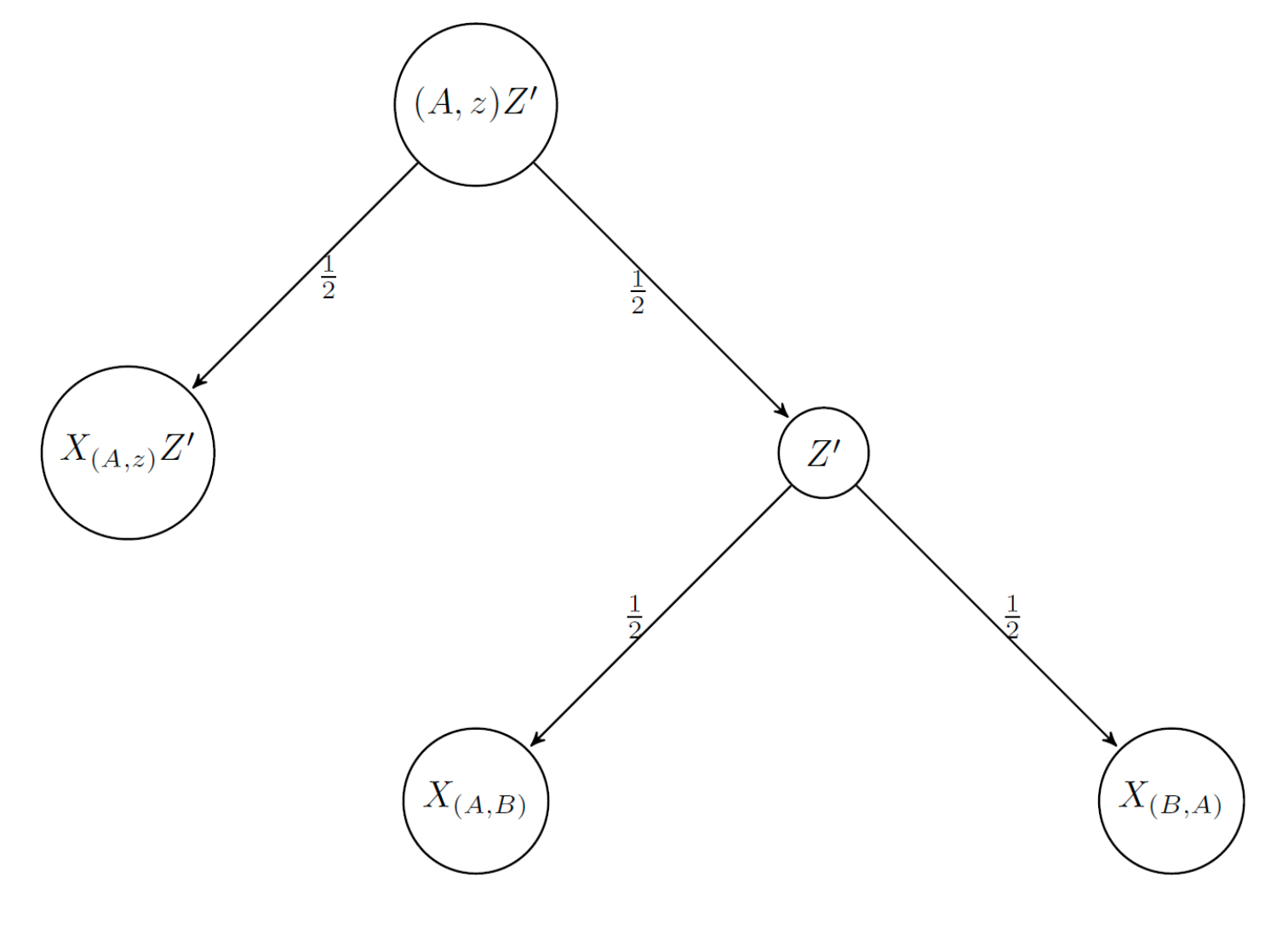}}
\caption{\label{fig:fig5} $(A,z)Z'$'s unfolding tree}
\end{figure}}
\end{enumerate}

\noindent Induction step: suppose the induction hypothesis for $l=n-1$ is true, i.e., $\mathcal{P}((x_2,y_2)(x_3,y_3)\cdots (x_n,y_n)Z',\varphi_1)=\rho({\rm trim}(x_2x_3\cdots x_n)Z')$.

Now we consider the case of $l=n$, i.e., $\mathcal{P}((x_1,y_1)\alpha' Z',\varphi_1)$ where $\alpha'=(x_2,y_2)\cdots (x_n,y_n)$. 

Note that $(x_1,y_1)\alpha'Z\rightarrow^{\frac{1}{2}}X_{(x_1,y_1)}\alpha'Z'\rightarrow^1\alpha' Z'$ and $(x_1,y_1)\alpha'Z'\rightarrow^{\frac{1}{2}}\alpha' Z'$, we have the following $3$ cases:
\begin{enumerate}
\item { if $x_1=\bullet$, then we have 
$$\aligned
\mathcal{P}((x_1,y_1)\alpha'Z',\varphi_1)=&\frac{1}{2}\rho({\rm trim}(x_2\cdots x_n)Z')\\
+&\frac{1}{2}\rho({\rm trim}(x_2\cdots x_n)Z')\\
=&\rho({\rm trim}(x_1x_2\cdots x_n)Z');
\endaligned$$
}
\item {if $x_1=B$, then we obtain 
$$\aligned
\mathcal{P}((x_1,y_1)\alpha' Z',\varphi_1)=&\frac{1}{2}\mathcal{P}(\alpha'Z',\varphi_1)\\
=&\frac{1}{2}\rho({\rm trim}(x_2\cdots x_n)Z')\\
=&\rho({\rm trim}(x_1x_2\cdots x_n)Z');
\endaligned$$
}
\item {if $x_1=A$, then we get 
$$\aligned
\mathcal{P}((x_1,y_1)\alpha'Z',\varphi_1)=&\frac{1}{2}+\frac{1}{2}\rho({\rm trim}(x_2\cdots x_n)Z')\\
=&\rho({\rm trim}(x_1x_2\cdots x_n)Z').
\endaligned$$
}
\end{enumerate}

From which it immediate follows that $\mathcal{P}(\{\pi\in\mbox{$Run(F\alpha Z')$}\,|\,\pi\models^{\nu}\varphi_1\})=\rho(u'_{j_1}u'_{j_2}\cdots u'_{j_k}Z')$. The similar arguments apply for $\mathcal{P}(\{\pi\in\mbox{$Run(S\alpha Z')$}\,|\,\pi\models^{\nu}\varphi_1\})=\overline{\rho}(v'_{j_1}v'_{j_2}\cdots v'_{j_k}Z')$.
\end{proof}

Now, the Theorem \ref{theorem1} can be proved naturally as follows:

\noindent{\em Proof of Theorem \ref{theorem1}}. Let $\pi$ be a path of pBPA $\triangle$, starting at $C$, induced by $C\alpha Z'$, where $\alpha$ is guessed by $\triangle$ as a solution of the modified PCP instance.

Then, we get
$$\aligned
(\ref{eq5})\text{ is }&\text{true}\\
&\quad\text{( by Lemma \ref{lemma3} )}\\
\Leftrightarrow& \,\,\mathcal{M}_{\triangle}, N\alpha Z'\models^{\nu}\mathcal{P}_{=\frac{t}{2}}(\varphi_1)\wedge\mathcal{P}_{\frac{1-t}{2}}(\varphi_2)\\
&\quad\text{( by $C\rightarrow N$ )}\\
\Leftrightarrow&\,\,\mathcal{M}_{\triangle},C\alpha Z\models^{\nu}{\bf X}[\mathcal{P}_{=\frac{t}{2}}(\varphi_1)\wedge\mathcal{P}_{=\frac{1-t}{2}}(\varphi_2)]\\
&\quad\text{( by $\mathcal{P}(C\rightarrow N)=1$ )}\\
\Leftrightarrow&\,\,\mathcal{M}_{\triangle},C\models^{\nu}\mathcal{P}_{=1}({\bf X}[\mathcal{P}_{=\frac{t}{2}}(\varphi_1)\wedge\mathcal{P}_{=\frac{1-t}{2}}(\varphi_2)])\\
&\quad\text{( by Lemma \ref{lemma1} )}\\
\Leftrightarrow&\,\,\mathcal{M}_{\triangle},Z\models^{\nu}\mathcal{P}_{>0}({\bf true}{\bf U}[C\wedge \mathcal{P}_{=1}({\bf X}[\mathcal{P}_{=\frac{t}{2}}(\varphi_1)\wedge\mathcal{P}_{=\frac{1-t}{2}}(\varphi_2)])])
\endaligned$$

Thus
\begin{equation}
\label{eq9}
\begin{split}
\mathcal{M}_{\triangle},Z\models^{\nu}\mathcal{P}_{>0}({\bf true}{\bf U}[C\wedge \mathcal{P}_{=1}({\bf X}[\mathcal{P}_{=\frac{t}{2}}(\varphi_1)\wedge\mathcal{P}_{=\frac{1-t}{2}}(\varphi_2)])])
\end{split}
\end{equation}

if and only if $\alpha$ is a solution of the modified PCP instance. As a result, an algorithm for determining whether (\ref{eq9}) is true contributes to an algorithm for solving the modified Post Correspondence Problem.\Q.E.D

\begin{remark}
\label{remark7}
Some may argue that the PCTL formula given in (\ref{eq9}) is not well-formed, since it contains ``parameter" $t$. In fact, $t$ should be viewed as a rational constant. To see so, let us consider the following well-formed PCTL formula which contains no ``parameter" $t$:
\begin{eqnarray}
\label{eq10}
\mathcal{P}_{>0}({\bf true}{\bf U}[C\wedge \mathcal{P}_{=1}({\bf X}[\mathcal{P}_{=\frac{1}{6}}(\varphi_1)\wedge\mathcal{P}_{=\frac{1}{3}}(\varphi_2)])])
\end{eqnarray}
Now, (\ref{eq10}) is well-formed and it is not hard to see that $\rho(u'_{j_1}\cdots u'_{j_k}Z')=\frac{1}{3}$ and $\overline{\rho}(v'_{j_1}\cdots v'_{j_k}Z')=1-\frac{1}{3}=\frac{2}{3}$. Namely, just let $t=\frac{1}{3}$.

It meets the following condition: $\rho(u'_{j_1}\cdots u'_{j_k}Z')+\overline{\rho}(v'_{j_1}\cdots v'_{j_k}Z')=1$. By Lemma \ref{lemma2}, one has that $u'_{j_1}u'_{j_2}\cdots u'_{j_k} = v'_{j_1}v'_{j_2}\cdots v'_{j_k}$. So, an algorithm for checking whether (\ref{eq10}) is true will lead to an algorithm to solve the modified PCP problem. \textcolor{red}{Some reader cannot see how the problem of evaluating the formula for all possible values of the parameter, which is infinite, could be overcome. For this, since our topic is undecidability of the issue, it is enough for us to find a well-formed formula. So how to find all possible values of the parameter, which is infinite, is out of our topic.}
\end{remark}

\begin{remark}
\label{remark8}
Although \cite{BBFK14} has reached the result that $\mathcal{P}(N\alpha Z',\varphi_1\vee \varphi_2) =1$, \cite{BBFK14} was unable to construct the PCTL formula (\ref{eq10}) based on the above relation deduced by themselves. Of course, our approach in fact is also based on this relation and the difference is that we are able to translate the above relation to $\mathcal{P}_{=\frac{t}{2}}(\varphi_1)\wedge\mathcal{P}_{=\frac{1-t}{2}}(\varphi_2)$, which is crucial to constructing the PCTL formula (\ref{eq10}).
\end{remark}

\begin{remark}
\label{remark9}
In fact, we can add a finite number of $N_i$ to the stack alphabet $\Gamma$, as well as a sufficient number of rules $C \rightarrow N_1 \rightarrow N_2 \rightarrow \cdots\rightarrow N_k\rightarrow N$ to $\delta$. Hence, the PCTL formula $\mathcal{P}_{>0}({\bf true}{\bf U}[C\wedge\mathcal{P}_{=1}({\bf true}{\bf U}\mathcal{P}_{=1}[{\bf X}(\mathcal{P}_{=\frac{t}{2}}(\varphi_1)\wedge\mathcal{P}_{=\frac{1-t}{2}}(\varphi_2))])])$ is also valid.

Furthermore, if we change the transition rule from $C\rightarrow N$ to $C\rightarrow F\text{ $|$ }S$, the formula $\mathcal{P}_{>0}({\bf true}{\bf U}[C\wedge\mathcal{P}_{=\frac{t}{2}}(\varphi_1)\wedge\mathcal{P}_{=\frac{1-t}{2}}(\varphi_2)])$ is much simpler. \textcolor{red}{Note that changing the transition rule from $C\rightarrow N$ to $C\rightarrow F\,|\,S$ corresponds to the formula: $\mathcal{P}_{>0}({\bf true}{\bf U}[C\wedge\mathcal{P}_{=\frac{t}{2}}(\varphi_1)\wedge\mathcal{P}_{=\frac{1-t}{2}}(\varphi_2)])$. But the proof should also be changed.}
\end{remark}

\section{Conclusions}
\label{sec:conclusion}

In this paper we have shown that the model-checking question for {\em stateless probabilistic pushdown systems} against the PCTL is generally undecidable, herein settling a common open question in \cite{EKM06,BBFK14}. We should point out that, our work can be seen as a continuation of the pioneering works \cite{EKM06,BBFK14}.

\section*{Acknowledgments}
\label{sec:acknowledgements}
Sincere thanks from the second author go to Dr. Forejt  \cite{For13} for answering our questions about the modified PCP. Furthermore, anonymous readers provided many invaluable suggestions for us to improve the manuscript.



\end{document}